\documentclass[a4paper,12pt]{article}
\pdfoutput=1 

\usepackage{jcappub} 
\usepackage{natbib}
\usepackage[T1]{fontenc}
\title{Updated Constraints on Primordial Black Hole Evaporation}

\author[a]{Mrunal Korwar,}
\author[b]{Stefano Profumo}

\affiliation[a]{Department of Physics, University of Wisconsin-Madison, Madison, WI 53706, USA}
\affiliation[b]{Department of Physics and Santa Cruz Institute for Particle Physics,\\
University of California, Santa Cruz, CA 95064, USA}

\emailAdd{mkorwar@wisc.edu}
\emailAdd{profumo@ucsc.edu}

\abstract{The Hawking evaporation process, leading to the production of detectable particle species, constrains the abundance of light black holes, presumably of primordial origin. Here, we reconsider and correct constraints from soft gamma-ray observations, including of the gamma-ray line, at 511 keV, produced by electron-positron pair-annihilation, where positrons originate from black hole evaporation. First, we point out that the INTEGRAL detection of the Large Magellanic Cloud provides one of the strongest bounds attainable with present observations; and that future MeV gamma-ray telescopes, such as GECCO, will greatly enhance such constraints. Second, we discuss issues with previous limits from the isotropic flux at 511 keV and we provide updated, robust constraints from recent measurements of the diffuse Galactic soft gamma-ray emission and from the isotropic soft gamma-ray background.
}

\begin{document} 
\maketitle
\flushbottom

\section{Introduction}
\label{sec:intro}
The persistent absence of any definitive, non-gravitational signal from dark matter (DM) candidates \cite{ParticleDataGroup:2022pth}, including those in the class of weakly interacting massive particles \cite{Arcadi:2017kky}, has led to a renewed interest in black holes of non-stellar origin, also known as primordial black holes (PBHs) \cite{Carr:2020gox, Auffinger:2022khh, Escriva:2022duf}. While at sufficiently large masses gravitational (micro-)lensing strongly constrains  the fraction of the cosmological DM that can consist of PBHs \cite{Carr:2020gox}, finite-size source effects make it virtually impossible to probe masses significantly below around 10$^{20}$ grams \cite{Smyth:2019whb}. Nonetheless, around that mass black holes are slated to emit abundant and bright non-thermal radiation via Hawking evaporation,  offering an alternative way of constraining their global contribution to the cosmological DM \cite{Auffinger:2022khh}.

The temperature associated with black holes is inversely proportional to the hole's mass, and $T_{\rm BH}\sim 0.1$ keV $(10^{20}$ grams$/M_{\rm BH})$, thus the emitted radiation, for masses below around 10$^{20}$ grams is expected in the X-ray to gamma-ray energy range; additionally, if $M_{\rm BH}\lesssim 10^{16}$ grams, evaporation also copiously produces positrons which, in turn, upon annihilating with ambient electrons, give rise to detectable gamma rays, in particular at an energy $E_\gamma=m_e=511$ keV \cite{MacGibbon:1991vc}.

This observation has led to attempts at constraining the mass fraction of black holes around $M_{\rm BH}\sim 10^{16}$ grams, or possibly even heavier, using observations at 511 keV \cite{MacGibbon:1991vc, Bambi:2008kx, Carr:2009jm}; in particular, the recent Ref.~\cite{DeRocco:2019fjq} obtained somewhat conservative limits, using the observed 511 keV emission from the Galactic center,  for a Navarro-Frenk-White Galactic DM density profile \cite{Navarro:1995iw}, and for log-normal mass distribution for the Galactic PBHs. 

Additionally, recent analyses have explored constraints from the isotropic hard X-ray background, stemming from both the 511 keV photons from positron annihilation as well as from photons directly produced from evaporation (e.g. \cite{Iguaz:2021irx}), as well as utilizing template fits to the SPectrometer aboard the INTEGRAL (SPI) satellite to derive constraints on the diffuse Galactic emission associated with decaying DM \cite{Berteaud:2022tws}, albeit, in this latter case, without including the emission from positron annihilation.

Here, we reassess limits from the 511 keV emission from positron annihilation and from direct photon production from Hawking evaporation, and discuss new targets and issues with certain data sets and methods considered in the recent past. The reminder of this note is structured as follows: In sec.~\ref{sec:PBHevap}, we discuss the details of the computation of positron production from Hawking evaporation, as well as positrons transport and annihilation; in the following sec.~\ref{sec:galsource} we give details on the positron flux produced in different directions in the sky, including extragalactic sources and the Galactic center; sec.~\ref{sec:egalsource} discusses limits from the measurement of the isotropic hard X-ray background, and the final sec.~\ref{sec:disc} presents our discussion and conclusions.

\section{PBH Evaporation and Positron Production}
\label{sec:PBHevap}
A non-rotating Black Hole (BH) with mass $M$ has a temperature of \cite{Hawking:1975vcx}
\begin{equation}
T_{\rm BH} = \frac{1}{8\pi G_{N} M} \approx 1.06 \, {\rm MeV}\ \left(\frac{10^{16}\ {\rm  gm}}{M}\right),
%
\end{equation}
where $G_{N}$ denotes Newtons gravitational constant.  Such BH radiates each fundamental particle species at a rate (in natural units $\hbar=c=1$)
\begin{equation}\label{eq:emissionrate}
\frac{\partial^{2}N_{i}}{\partial E_{i}\partial t}= \frac{g_{i}}{2\pi} \frac{\Gamma_{i}(T_{\rm{BH}},E_{i},s_{i})}{\exp(E_{i}/T_{\rm BH})-(-1)^{2s_{i}}} \, ,
\end{equation}
where $g_{i}$ denotes the degrees of freedom of species $i$, with $g_{e}=2$ (positrons) and $g_{\gamma}=2$. The 
$\Gamma_{i}$ are species-dependent greybody factors, with $E_{i}$ indicating the energy of the emitted particle, and $s_i$ its spin. The greybody factors are typically written in the form 
\begin{equation}
\Gamma_{i}(E_{i},T_{\rm BH},s_{i})= 27 \Big{(}\frac{E_{i}}{8\pi T_{\rm BH}}\Big{)}^{2} \gamma(E_{i}, T_{\rm BH}, s_{i}) \, ,
\end{equation}
with $\gamma(E_{i}, T_{\rm BH}, s_{i})$ factors that tend to 1 at $E_{i}\gg T_{\rm BH}$, while $\gamma \ll 1$ at $E<T_{\rm BH}$. We use $\gamma$ factors from  \,\cite{Ukwatta:2015iba}, which agrees with the BlackHawk code \cite{Arbey:2019mbc}). 

In this paper, we are interested in primary and secondary emission from PBH with mass $10^{16}\, \rm gm\lesssim M\lesssim 10^{18}\, \rm gm$; since $T_{\rm BH}< \rm MeV$ in this mass range, the emission we focus on lies in the soft gamma-ray and hard X-ray bands. In addition to the primary photon emission from PBH, we also consider the emission from positrons annihilating with the thermal electrons in the Interstellar medium (ISM), leading to a 511 keV line and to the so-called ortho-positronium continuum. For the latter, we note that while the primary electrons/positrons emitted from PBH would be relativistic initially, they lose energy via Compton scattering and ionization losses and turn non-relativistic well before annihilation (albeit a small contribution from annihilation in flight is possible: the fraction of positrons that annihilate before turning non-relativistic is typically less than 5\% for the initial energy of electrons up to 10 MeV \cite{Prantzos:2010wi, Keith:2021guq}). 
Qualitatively this can be understood considering the energy loss and annihilation timescales of relativistic electrons in the typical ISM condition with $n_{e}\approx 1 \, \rm cm^{-3}$:
The energy loss timescale for a 10 MeV electron is $\tau_{\rm loss} \approx E/|dE/dt|_{\rm loss} \approx 10^{13}\,\rm s$, while the annihilation timescale is $\tau_{\rm ann}= (\sigma_{\rm ann} v n_{e})^{-1} \approx 10^{15} \, \rm s$ . 
This implies $\tau_{\rm loss}/\tau_{\rm ann}\ll1$, independent of $n_e$, since both time scales have the same dependence on $n_e$,  justifying the assumption we make hereafter that  positrons turn non-relativistic before annihilation. We also note that the typical diffusion length-scale for positrons, $d_{\rm diff}\approx (D(E) \tau_{\rm loss})^{1/2}$, is typically small for positrons: for typical values of $D(E)\approx 3\times 10^{28} (E/\rm GeV)^{-1/3}\rm cm^{2} \rm s^{-1}$ we have $d_{\rm diff}\approx 0.1 \, \rm kpc$ for positrons with energies of 10 MeV (note that in what follows we generally expect positrons to be dominantly at even lower energies, with resulting even smaller diffusion lengths). While uncertainties exist on both the value of $D(E)$ within the Galactic diffusive halo, and even greater ones on the values pertinent for extragalactic systems, we estimate that net diffusion of positrons prior to annihilation outside our targets of interest to be a negligible effect. 
The determination of the diffusion coefficient depends on a variety of assumptions on the diffusion model, and is an area of ongoing active research. The value we assume is commonly taken as the central value based on comparisons of a number of numerical codes modeling cosmic-ray transport with observational data on cosmic-ray species \cite{Strong:2007nh, Genolini:2015cta}. While deriving precise uncertainty brackets to the diffusion coefficient itself absent assumptions about other transport model parameters is problematic, such uncertainty can be solidly considered to be within a factor of a few at most of the central value, based on the results of the studies presented, e.g., in \cite{putze2010markov,di2010unified,Trotta:2010mx}.

\section{Constraints from the Galactic center, M31, and the Large Magellanic Cloud}
\label{sec:galsource}
In this section, we consider limits on the PBH DM fraction derived using INTEGRAL/SPI data. Generically, the flux of 511 keV photons from PBH evaporation in a region of solid angle $\Delta \Omega$ is given by
\begin{align}
\Phi_{511} &= \frac{\mathcal{L}_{511}(M) f_{\rm PBH}}{4\pi M} \,\int_{\Delta \Omega}D(\Omega) d\Omega \nonumber \, ,\\
&= \frac{f_{\rm{line}}f_{\rm PBH}}{4\pi M}\frac{dN_{e}}{dt}\,\int_{\Delta \Omega}\int_{\rm{l.o.s}}\rho_{\rm DM} dr d\Omega \,.
\end{align}
The symbol $f_{\rm PBH}$, above and thereafter, indicates the fraction of the cosmological DM that is allowed to consist of PBH with a single mass $M$. Also in the equation above, $\mathcal{L}_{511}(M)$ is the rate of 511 keV photons produced by a PBH of mass $M$, which, in turn, is related to the energy-integrated rate of positron emission $d^{2}N_{e}/dE dt$ and to the number of 511 keV photons produced when positrons interact with the thermal electrons, $f_{\rm line}$. Explicitly we have
\begin{equation}
  \mathcal{L}_{511}= f_{\rm line} \frac{dN_{e}}{dt}=   f_{\rm line} \int_{m_{e}}^{\infty}\, dE \, \frac{d^{2}N_{e}}{dE dt}\, .  
\end{equation}

The annihilating positrons could either directly produce two quasi-monochromatic photons of energy $E_\gamma\sim m_e$, or they could form a Positronium bound state with an electron, which, in turn, results $25\%$ of the time in two 511 keV photons, and $75\%$ of the time in three photons, each with energy $E_{\gamma}< 511 \, \rm keV$. Thus,  numerically one has $f_{\rm line}=2(1-f_{\rm p}) + 2 f_{\rm p}/4$, where $f_{\rm p}$ is the Positronium fraction. For the interstellar medium of the Milky Way $f_{\rm p}=0.967\pm 0.022$ \cite{Jean:2005af} which results in $f_{\rm line}\approx 0.55$, while, more generally, or absent a direct observational determination of $f_{\rm line}$,  $0.5 \leq f_{\rm line} \leq 2$.

The total flux additionally depends on the $D(\Omega)$-factor which is, as customary, the angular-averaged integral over the line of sight of the DM density i.e.$D(\Omega) = \int_{\rm{l.o.s}} \rho_{\rm DM} dr$. For the DM distribution, we consider a Navarro-Frenk-White (NFW) halo profile \cite{Navarro:1995iw}
\begin{equation}
\rho_{\rm DM} = \frac{\rho_{0} r_{s}^{3}}{r (r_{s}+r)^{2}} \, ,
\end{equation}
where the parameters $\rho_{0},r_{s}$ are system dependent. Notice that we numerically evaluated  the impact of utilizing alternate DM density profiles, such as the cored Burkert \cite{Nesti:2013uwa} or the cuspier Einasto profile \cite{1965TrAlm...5...87E, 2019JCAP...10..037D} on the $D(\Omega)$-factor and found that it is less than a 5\% effect.

\begin{figure}
    \centering
    \includegraphics[width=0.75\textwidth]{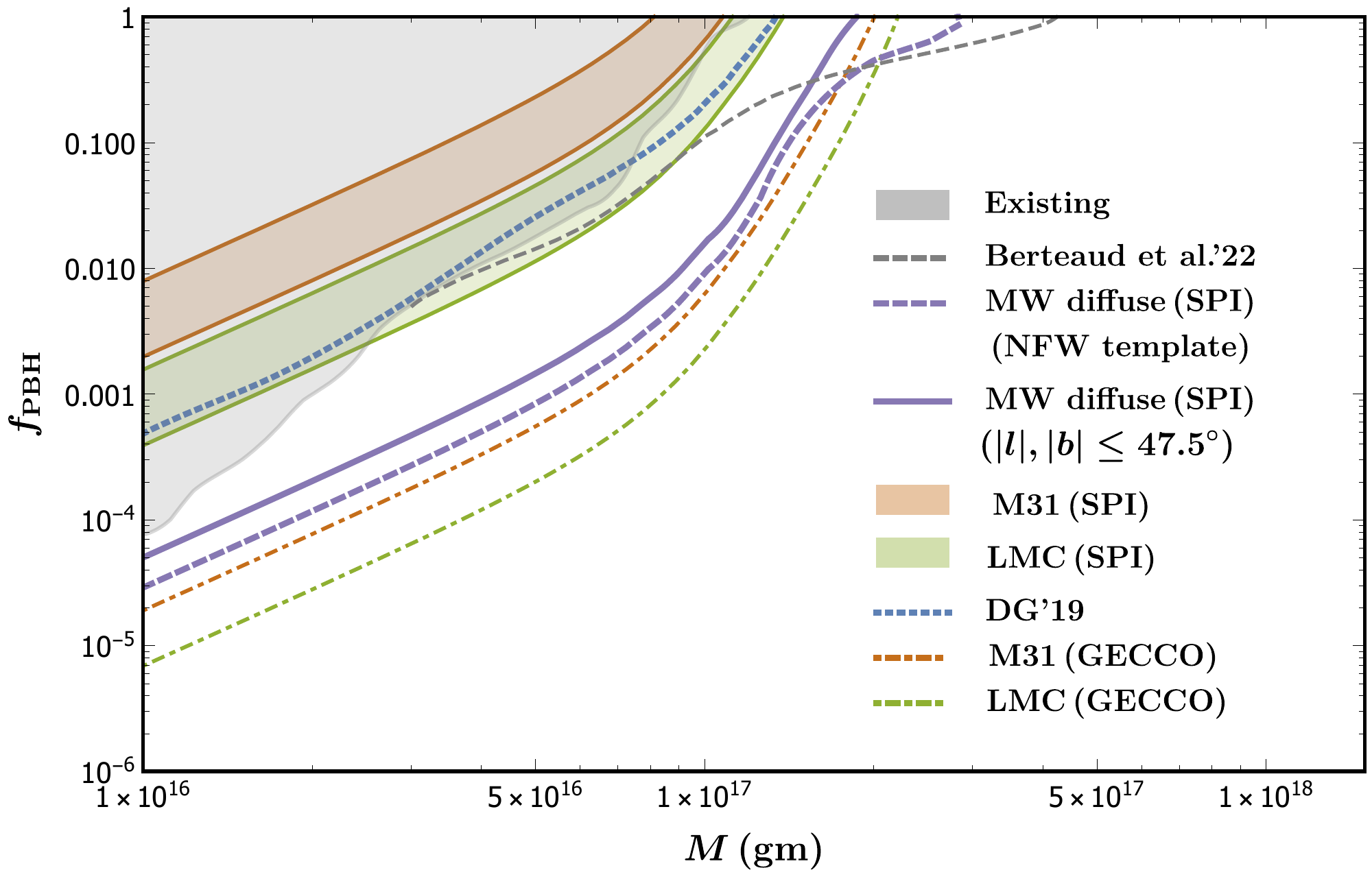}
    \caption{Bounds on the PBH DM fraction as a function of mass from MW diffuse emission of 511 keV line and soft gamma-ray continuum in the region ($|l| \leq 47.5^{\circ}$, $|b| \leq 47.5^{\circ}$) as observed by SPI \cite{Siegert:2021trw, Berteaud:2022tws} (purple line), and using extracted spectra utilizing NFW template \cite{Berteaud:2022tws}(dashed purple line); upper limits on the flux of 511 keV line from LMC \cite{Siegert:2015knp} (green band) and M31 \cite{Bandyopadhyay:2008ts} (brown band) from SPI upper limits. The band considers the uncertainty in positron fraction $f_{\rm p}$ in these two sources.
    We show the GECCO \cite{Coogan:2021rez} sensitivity in probing the 511 keV line from the LMC (green dot-dashed) and M31 (brown dot-dashed), assuming a conservative value of $f_{\rm line}=0.5$ and using worst-case sensitivity for point sources. We also show the limits derived by DeRocco and Graham'19 (DG'19) (blue dotted line) using the observation of the 511 keV line from the Galactic center  \cite{DeRocco:2019fjq}. The grey dotted line shows the bounds from \cite{Berteaud:2022tws} which uses template fit method on the Galactic emission data \cite{Siegert:2021trw, Berteaud:2022tws} to derive the bound.
    The grey band shows the existing constraints on PBH in the shown mass range \cite{Coogan:2020tuf}.}
    \label{fig:gal_pbh}
\end{figure}

The SPI spectrometer on the INTEGRAL satellite has observed the flux of 511 keV photons $(1.07\pm 0.03)\times 10^{-3} \, \rm photons \, \rm cm^{-2}\, \rm s^{-1}$ from the inner regions of Milky Way \cite{Weidenspointner:2007rs}. This has been used to constrain the fraction of DM in PBH in Ref.~\cite{DeRocco:2019fjq}, which obtained  limits we show in Fig.\ref{fig:gal_pbh} with a dotted blue line assuming, conservatively, that the flux from PBH evaporation cannot exceed the {\sl total} observed flux. Recently, a new analysis of a larger data set of 16-years of INTEGRAL/SPI observations containing diffuse soft gamma-ray emission from the region ($|l|,|b|\leq 47.5^{\circ}$) was analyzed \cite{Siegert:2021trw, Berteaud:2022tws}. Utilizing a template fit technique, Ref.~\cite{Berteaud:2022tws} extracts the amplitude of the signal associated with a putative decaying DM species which, in turn, can be associated here with the  emission from PBH evaporation. Here, we use the results of Ref.~\cite{Berteaud:2022tws} in two ways: first, we conservatively posit that the emission from PBH evaporation does not exceed the total emission in the region of interest (i.e. the left panel of fig.~2 of Ref.~\cite{Berteaud:2022tws}) (shown in Fig.\ref{fig:gal_pbh} by a purple line); secondly, we demand that the emission from PBH evaporation in the Galactic region of interest does not exceed the intensity associated with the  line-of-sight integrated NFW profile, i.e. the right panel of fig.~2 of Ref.~\cite{Berteaud:2022tws} (shown in Fig.\ref{fig:gal_pbh} by a dashed purple line). Note that this latter approach assumes that positron propagation prior to annihilation does not significantly affect the 511 keV morphology which, as we claimed in the previous section, is a reasonable assumption given the relevant time-scales. Notice that the bound shown by the purple and dashed purple line in Fig.\ref{fig:gal_pbh} is stronger compared to the one derived in \cite{Berteaud:2022tws} (dashed gray line), which uses the the same data. This is because Ref.\cite{Berteaud:2022tws} neglected the contribution to the diffused soft gamma-rays from positron annihilation, which we find dominates the signal.

If DM consists at least in part of PBH, then the  511 keV line resulting from PBH positrons produced in PBH evaporation should exist in any DM halo, provided electrons are present; as such, potentially promising targets for the detection of a bright 511 keV signal include the Large Magellanic Cloud (LMC) and Andromeda galaxy (M31). 
Neither target was detected by INTEGRAL/SPI, providing, as such, promising opportunities to constrain the mass fraction of PBH relative to DM $f_{\rm PBH}$. Using the all-sky map of 511 keV photons by INTEGRAL/SPI \cite{Bandyopadhyay:2008ts}, the upper limit on the flux of 511 keV photons has been placed at $3.6 \times 10^{-5} \rm cm^{-2} \rm s^{-1}$ and $10^{-4} \rm cm^{-2} \rm s^{-1}$ for LMC \cite{Siegert:2015knp} and M31 \cite{Bandyopadhyay:2008ts}, respectively. Since we lack direct observational data for $f_{\rm p}$ for the two targets, we show the band between the upper and lower limits derived using $f_{\rm line}=0.5$ and $f_{\rm line}=2$ . The parameters for the NFW profile are taken from Ref.~\cite{Buckley:2015doa} for the LMC and from Ref.~\cite{Geehan:2005aq} for M31. We associate the solid angle of the observation $\Delta \Omega \approx 3^{\circ}$ with the imaging resolution of the SPI instrument \cite{2003A&A...411L..63V}. We show the limits from LMC and M31 in Fig.~\ref{fig:gal_pbh} by green and brown band, respectively, and note that the LMC limits exceed other existing limits, shown in grey, taken from Ref.~\cite{Coogan:2020tuf} (to which we refer the Reader for details). Note that to derive the limits shown in this section we do not include here the isotropic extragalactic diffuse emission from evaporating PBH (see next section).

Note that the the limits on the 511 keV flux from the LMC and M31 are limited by the the sensitivity of the SPI spectrometer. Future telescopes such GECCO \cite{Orlando:2021get}, whose science program includes specifically investigating the origin of the 511 keV emission in the Galactic center, will conduct high-sensitivity measurements of cosmic $\gamma$-radiation in the energy range from 100 \rm keV-10 \rm MeV.  The GECCO instrument will have a field of view of $\sim 5-6^{\circ}$, and a narrow-line sensitivity, in the best case scenario, of $7.4\times 10^{-8}\,\rm cm^{-2} \rm s^{-1}$, and in the worse case scenario $3.2\times 10^{-7}\,\rm cm^{-2} \rm s^{-1}$ for point sources \cite{Coogan:2021rez}. 

While estimating a bound using  sensitivities alone would require background modelling of the 511 keV line from these sources, which is at present not available because of the lack of the observation of the line from these sources with current telescopes. In Fig. \ref{fig:gal_pbh} we thus show the sensitivity curves assuming no background 511 keV line will be detected by GECCO. We take the conservative value of $f_{\rm line}=0.5$ and use the worst case scenario sensitivity for point sources. We see that GECCO will be able to probe PBHs with much smaller DM fraction below few times $10^{17}\, \rm gm$. The GECCO discovery potential for M31 and the LMC is comparable to the shown constraining power.

\section{Constraints from the Isotropic X-ray background}
\label{sec:egalsource}
In this section, we consider measurements of the {\sl isotropic} X-ray and soft gamma-ray emission to constrain the PBH DM abundance. There are two main contributions to the isotropic flux from PBH: (1) direct photon radiation from PBH evaporation and (2) positrons annihilating with thermal electrons producing an X-ray background at energies around and below 511 keV. 
The diffuse photon emission consists of a contribution from PBH from {\em extragalactic}  structures at all redshifts, given by
\begin{equation}\label{eq:egalgamma}
\frac{dF_{\gamma}^{\rm egal}(E)}{dE} = \frac{f_{\rm PBH}\Omega_{\rm DM}\rho_{\rm c}}{4\pi M}\int_{0}^{\rm z_{\rm max}}\frac{dz}{H(z)}\frac{d^{2}N_{\gamma}}{dEdt}(E(1+z))\, ;
\end{equation}
in the equation above, $f_{\rm PBH}$ is the fraction of DM in PBH, $\rho_{c}=9.1 \times 10^{-30} \rm gm \,\rm cm^{-3}$, $H(z)=H_{0}\sqrt{\Omega_{\Lambda} + (1+z)^{3} \Omega_{m}}$, $H_{0}=67.36 \, \rm km\, \rm s^{-1} \rm Mpc^{-1}$, $z_{\rm max}=100$, $\Omega_{\Lambda}=0.6847$, $\Omega_{\rm DM}=0.2645$ \cite{Planck:2018vyg}. The second contribution to the diffuse emission stems from PBHs inside our own Galaxy; this contribution, while not isotropic, contributes, at its lowest level on the sky, to the isotropic emission as well. The Galactic diffuse emission from direct evaporation is given by
\begin{equation}\label{eq:galgamma}
\frac{dF_{\gamma}^{\rm gal}(\Omega)}{dE} = \frac{f_{\rm PBH} D(\Omega)}{4\pi M}\frac{d^{2}N_{\gamma}}{dE dt} \, .
\end{equation}
Above, $D(\Omega)$, as in the previous section, is the angular average of the line-of-sight integral of the dark matter density, which depends, here, on the direction of observation in Galactic coordinates $(l,b)$. In earlier work \cite{Iguaz:2021irx} that sought to constrain PBH evaporation using the isotropic background, the authors utilized the direction, in Eq.~(\ref{eq:galgamma}) above, $l=180^{\circ}, b=0$ i.e. in the anti-Galactic center. This is a conservative choice, since $D(\Omega)$ is at its minimum for the anti-Galactic center direction; however, this choice is actually overly conservative:  measurements of the diffuse Galactic hard X-ray/soft-$\gamma$-ray emission indicate that at a Galactic latitude of $b=15^{\circ}$, the diffuse emission from the Galaxy is highly suppressed \cite{Siegert:2022jii}. Seeking to optimize the line of sight integral $D(\Omega)$, we thus choose the $l=0,b=15^{\circ}$ direction, which has negligible Galactic diffuse flux contamination to set out limits from the isotropic flux; notice that unlike in Ref.~\cite{Berteaud:2022tws}, that utilizes a template fit analysis to subtract off a number of different background contributions, here we conservatively limit the PBH emission by requiring it to be lower than the observed {\sl total} flux. 

Next, we turn to positron emission by PBH and their annihilation, which also contributes to the isotropic X-ray background. In this case, we again have to take into account the uncertainty in the $f_{\rm p}$. We consider two cases $f_{\rm p}=0$ and $f_{\rm p}=1$. The Galactic contribution is given by
\begin{align}
\frac{dF_{0}^{\rm gal}(\Omega)}{dE} &= \frac{f_{\rm PBH} D(\Omega)}{4\pi M}\frac{dN_{e}}{dt} 2 \delta(E-m_{e}) \, , \label{eq:fp0gal}\\ 
\frac{dF_{1}^{\rm gal}(\Omega)}{dE} &= \frac{f_{\rm PBH} D(\Omega)}{4\pi M}\frac{dN_{e}}{dt} \Big{(}\frac{2}{4} \delta(E-m_{e}) + \frac{9}{4} \frac{h_{3\gamma}(E/m_{e})}{m_{e}}\Big{)} \, .\label{eq:fp1gal}
\end{align}
For $f_{\rm p}=0$, the annihilation of electrons and positrons leads to 2 photons contributing to the isotropic 511 keV flux. For the $f_{\rm p}=1$ case, annihilation produces, in addition to the 511 keV line, a continuum spectrum $h_{3\gamma}(E/m_{e})$, up to $E=m_{e}$ and is given by \cite{Ore:1949te,Manohar:2003xv}
\begin{equation}
h_{3\gamma}(x) = \frac{2}{\pi^{2}-9}\Big{[}\frac{2-x}{x} + \frac{(1-x)x}{(2-x)^{2}} - \Bigg{(}\frac{2(1-x)^{2}}{(2-x)^{3}} - \frac{2(1-x)}{x^{2}}\Bigg{)}\log(1-x)\Big{]}\, ,
\end{equation}
with $x=E/m_{e}$. Note that $\int_{0}^{1}\, dx h_{3\gamma}(x)=1$, and the integrated energy flux is  $\int \, dE \, E \,dF_{0}^{\rm gal}/dE = \int dE \, E \,dF_{1}^{\rm gal}/dE \propto 2 m_{e}$. 

As far as the extragalactic contribution to the isotropic X-ray flux from positron annihilation is concerned, we again consider the case of $f_{\rm p}=0$ and $f_{\rm p}=1$; for $f_{\rm p}=0$ the contribution is 
\begin{align}\label{eq:fp0egal}
\frac{dF_{0}^{\rm egal}}{dE} &= \frac{f_{\rm PBH}\Omega_{\rm DM}\rho_{c}}{4\pi M}\frac{dN_{e}}{dt} \int_{0}^{\rm z_{\rm max}}\,\frac{dz}{H(z)}\mathcal{F}(z) \,2\, \delta(E(1+z)-m_{e}) \, , \nonumber\\
&= \frac{f_{\rm PBH}\Omega_{\rm DM}\rho_{c}}{4\pi M}\frac{dN_{e}}{dt} \frac{2}{E} \frac{\, \theta(m_{e}-E)}{ H(m_{e}/E - 1)}\mathcal{F}(m_{e}/E - 1) \,,
\end{align}
where, in the equation above, $\theta(x)$ denotes the Heaviside step function. For $f_{p}=1$ we get
\begin{align}\label{eq:fp1egal}
\frac{dF_{1}^{\rm egal}}{dE} &= \frac{f_{\rm PBH}\Omega_{\rm DM}\rho_{c}}{4\pi M}\frac{dN_{e}}{dt} \int_{0}^{\rm z_{\rm max}}\,\frac{dz}{H(z)}\mathcal{F}(z) \,\Big{(}\frac{2}{4}\, \delta(E(1+z)-m_{e}) + \frac{9}{4} \frac{h_{3\gamma}(\frac{E(1+z)}{m_{e}})}{m_{e}} \Big{)}  \nonumber \, ,\\
&= \frac{f_{\rm PBH}\Omega_{\rm DM}\rho_{c}}{4\pi M}\frac{dN_{e}}{dt} \Big{[} \frac{2}{ 4 E} \frac{\, \theta(m_{e}-E)}{ H(m_{e}/E - 1)}\mathcal{F}(m_{e}/E - 1) + \nonumber \\  & \hspace{12em}\frac{9}{4 m_{e}} \int_{0}^{z_{\rm max}} dz\, \frac{h_{3\gamma}(\frac{E(1+z)}{m_{e}})}{H(z)}\mathcal{F}(z) \Big{]} \, .
\end{align}
Above, we have introduced the factor $\mathcal{F}(z)$, which takes into account the fraction of DM contained in DM halos which host galaxies and, thus, a dense, magnetized interstellar medium that confines the positrons and provide sufficiently large electron density for positron annihilation. The inclusion of the factor $\mathcal{F}(z)$ is necessary as the annihilation time scale of $e^{+}-e^{-}$ annihilation with IGM electron density of $n_{e}\approx 10^{-6} \rm cm^{-3}$ becomes larger than the age of the universe at lower redshifts. Whereas positrons released from PBH bound to DM halos hosting galaxies could annihilate with the electrons in the ISM medium where electron density is much larger. The theory of galaxy formation, along with the cosmological simulations, suggest that only a DM halo of mass greater than $M_{\rm min}\approx 10^{7}-10^{11}\, M_{\odot}$ could support the collapse of baryons to form galactic disks \cite{Rasera:2005sa, Rasera:2005gq}. 
The fraction of DM at redshift $z$ in halos with mass greater than $M_{\rm min}(z)$ is given by
\begin{equation}
\mathcal{F}(z) = \frac{1}{\rho_{m}(z)}\int_{M_{\rm min}(z)}^{\infty}dM \, M\,\frac{dn(M,z)}{dM} \, ,
\end{equation}
here $dn/dM$ is the halo mass function which by definition is normalized to $\int_{0}^{\infty}dM M \frac{dn}{dM}=\rho_{m}$ and thus $\mathcal{F}<1$ (for instance, we get $\mathcal{F}(0)\approx 0.5$)\cite{Cooray:2002dia, Ullio:2002pj}. We note that (1) this formalism is independent of the nature of the cold dark matter candidate under consideration (albeit for the masses under consideration, PBH and standard WIMP-like candidates do not differ in the halo formation history), and (2) that the $\mathcal{F}$ suppression factor has been neglected in earlier work \cite{Iguaz:2021irx}; thus, the inclusion of this factor produces a more robust bound\footnote{We also point out that Ref.~\cite{Iguaz:2021irx} over-estimates positron production by a factor of 2 due to a mis-interpretation of the output of the BlackHawk code \cite{Arbey:2019mbc}: The $g_{e}$ factor in eq.~(\ref{eq:emissionrate}) should be 2, for positron emission, whereas the BlackHawk code counts for both positron and electron emission to give $g_{e}=4$ (see Table III in \cite{Arbey:2019mbc}).}.  

\begin{figure}
    \centering
   \includegraphics[width=0.75\textwidth]{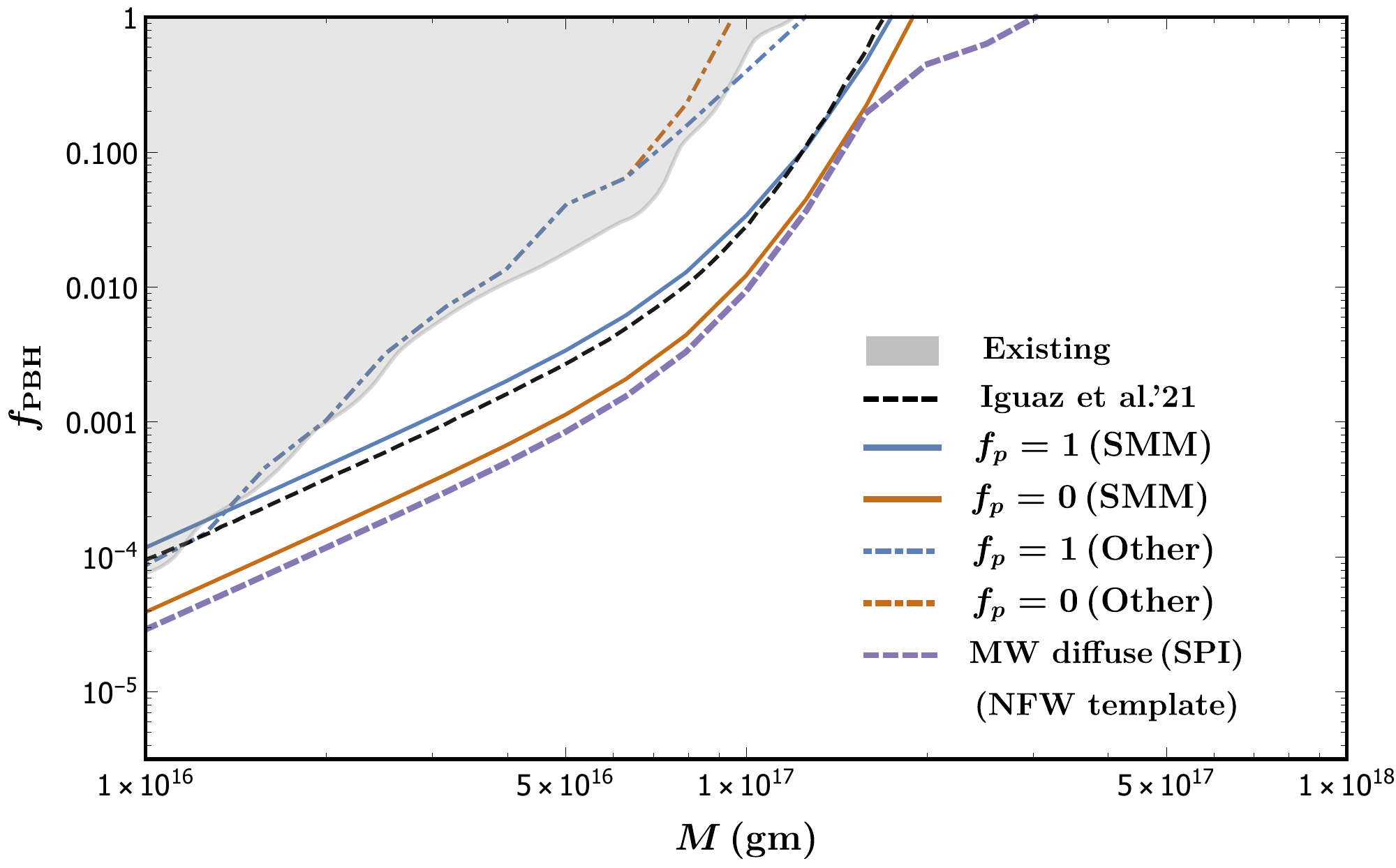}
    \caption{Bounds on the PBH DM fraction $f_{\rm PBH}$ as a function of mass from the observed isotropic X-ray and soft gamma-ray background flux. To take into account the uncertainty on the Positronium fraction, we show the bound obtained for both $f_{p}=0$ (brown) and $f_{p}=1$ (blue). The bounds are obtained by comparing the total flux in eq.~\ref{eq:total} to two different datasets; the first one is the SMM data set \cite{doi:10.1063/1.53933} (solid line), while the ``Other'' lines are derived using a combined data set that includes observations by COMPTEL \cite{doi:10.1063/1.1307028}, HEAO \cite{Kinzer_1997, Gruber:1999yr}, and BAT \cite{Ajello:2008xb} (dot-dashed line). For comparison, we show the strongest bound from Fig.\ref{fig:gal_pbh} shown here by the dashed purple line. We also show the strongest bound obtained in \cite{Iguaz:2021irx} (dashed black line, not corrected for an erroneous factor of 2 therein), which also considers isotropic X-ray flux data to constrain $f_{\rm PBH}$,
    and other existing constraints in this mass range are shown by gray shaded region \cite{Coogan:2020tuf}.}
    \label{fig:egal_pbh}
\end{figure}

The total isotropic X-ray contribution for $f_{p}=0$ and $f_{p}=1$ case is given by
\begin{align}\label{eq:total}
\frac{dF_{0/1}}{dE} &= \frac{dF_{\gamma}^{\rm egal}}{dE} + \frac{dF_{\gamma}^{\rm gal}}{dE} + \frac{dF_{0/1}^{\rm egal}}{dE} + \frac{dF_{0/1}^{\rm gal}}{dE} \, .
\end{align}
In order to bound $f_{\rm PBH}$ we compare $dF_{0/1}/dE$ with two different X-ray and soft gamma-ray datasets: the determination of the isotropic X-ray emission from the Solar Maximum Mission (SMM) \cite{doi:10.1063/1.53933}, and a combination of COMPTEL \cite{doi:10.1063/1.1307028} + HEAO \cite{Kinzer_1997, Gruber:1999yr} + BAT \cite{Ajello:2008xb}. Notice that the latter does not include data at 511 keV; thus, the limits only originate from direct photon emission and not from positron annihilation. We also marginalize over the delta function peak by considering finite energy resolution of $\Delta E/E \approx 7\,\%$ for the SMM instrument \cite{Share1997AcceleratedAA}. We obtain the bound by imposing that the X-ray flux due to PBH, eq.\ref{eq:total}, does not exceed any measurement data point by more than 2-$\sigma$. The bounds we obtain are shown in Fig.\ref{fig:egal_pbh}. 

While for the sake of comparison we show limits from the SMM measurement of the isotropic X-ray flux, we actually have strong reservations about how that measurement was derived and on the estimate of the error bars thereof. First, the data presented in Ref.~\cite{doi:10.1063/1.53933} were never peer-reviewed, nor is any detail on the analysis procedure offered in any publication. The analysis is presented in a PhD thesis\footnote{https://tigerprints.clemson.edu/arv\_dissertations/index.9.html}, which we reviewed and studied in detail. The thesis presents several outstanding issues, including frequent failure of the employed generalized non-linear least-squares fitting routine to converge and no robust evaluation of the systematic uncertainties other than a rough estimate  based on time fluctuations of the cosmic gamma-ray background, and the statement that ``{\sl an excellent detector, like SMM, has a resolution of about 7 percent}''; this latter issue is especially concerning since the measurement error is largely systematics-dominated. Finally, the claim that the measurement is compatible with that of the cosmic gamma-ray background (CGB) is exclusively based on weak evidence such as the listed items: ``{\sl (1) time variations of ``CGB'' term has no long term variations, no effects from the change of altitude, and no correlations with solar modulation, and (2) energy spectrum of ``CGB'' is quite different from any other background spectrum.}''. In view of the lack of consistent and peer-reviewed analysis presentation and of the deficiencies listed, we consider the SMM measurement as unreliable and advise against using it to set robust limits on new physics.

\section{Discussion and Summary}
\label{sec:disc}

We have clarified and re-assessed limits on the abundance of primordial black holes as dark matter candidates from gamma-ray primary and secondary emission resulting from Hawking evaporation of light, primordial black holes. Here, the secondary emission stems from the annihilation of positrons, produced by black holes' evaporation, with ambient electrons. We first considered new targets for which upper limits on the flux of photons at 511 keV from SPI observations are available, namely the Andromeda galaxy (M31) and the Large Magellanic Cloud; we showed that especially the latter offers highly competitive limits (albeit more uncertain due to the unknown local positron fraction), and re-evaluated the constraints from the diffuse gamma-ray emission in the direction of the Galactic center, which, at present, offer the best and most solid constraints on the largest-mass PBH-to-DM mass fraction. We then also re-evaluated constraints from measurements of the isotropic cosmic gamma-ray background and argued that limits from the SMM measurement thereof should not be considered robust, while other telescopes offer comparatively weaker limits than those from the Galactic diffuse emission.
Finally, we reiterated the point that future telescopes, such as the proposed GECCO telescope, will offer opportunities to discover Hawking evaporation, or constrain to a much higher degree of precision, the possibility that PBH forms a large fraction of the cosmological dark matter.

\acknowledgments
S.P. is partly supported by the U.S. Department of Energy grant number de-sc0010107. The work of MK is supported by the U.S. Department of Energy
under the contract DE-SC-0017647. We are very grateful to Pasquale Serpico for useful feedback and communications, and to Marco Ajello for communications and for kindly sharing with us the PhD thesis on SMM.

\setlength{\bibsep}{3pt}
\bibliographystyle{JHEP}
\bibliography{sample}

\end{document}